\pdfoutput=1
\documentclass[%
reprint,
superscriptaddress,
 amsmath,amssymb, 
 aps,
 prl,
floatfix,
]{revtex4-2}

\usepackage{graphicx}% Include figure files
\usepackage{epsfig} %Works faster than the graphicx package}
\usepackage{dcolumn}% Align table columns on decimal point
\usepackage{bm}% bold math
\usepackage{epstopdf}
\usepackage{multirow}
\usepackage{amsmath}
\usepackage{booktabs}
\usepackage{color}
\usepackage{gensymb}
\usepackage{kantlipsum}  
\usepackage{hyperref}
\usepackage{ulem}
\usepackage{braket}
\usepackage{physics}
\usepackage[utf8]{inputenc}
\usepackage[T1]{fontenc}
\usepackage{mathptmx}
\begin{document}

\title{Laser Scheme for Doppler Cooling of the Hydroxyl Cation (OH$^+$)}

\preprint{APS/123-QED}

\author{Niccol{\`o} Bigagli}
\affiliation{%
Department of Physics, Columbia University, New York, New York 10027, USA
}
\author{Daniel~W.~Savin}
\affiliation{%
Columbia Astrophysics Laboratory, Columbia University, New York, New York 10027, USA
}
\author{Sebastian~Will}\email{Corresponding author. Email: sebastian.will@columbia.edu}
\affiliation{%
Department of Physics, Columbia University, New York, New York 10027, USA
}

\date{\today}% It is always \today, today

\begin{abstract} 
We report on a cycling scheme for Doppler cooling of trapped OH$^+$ ions using transitions between the electronic ground state $X^3\Sigma^-$ and the first excited triplet state $A^3\Pi$. We have identified relevant transitions for photon cycling and repumping, have found that coupling into other electronic states is strongly suppressed, and have calculated the number of photon scatterings required to cool OH$^+$ to a temperature where Raman sideband cooling can take over. In contrast to the standard approach, where molecular ions are sympathetically cooled, our scheme does not require co-trapping of another species and opens the door to the creation of pure samples of cold molecular ions with potential applications in quantum information, quantum chemistry, and astrochemistry. The laser cooling scheme identified for OH$^+$ is efficient despite the absence of near-diagonal Franck-Condon factors, suggesting that broader classes of molecules and molecular ions are amenable to laser cooling than commonly assumed. 
\end{abstract}

\maketitle

\section{Introduction}

Laser cooling and quantum control of atoms and atomic ions has enabled a plethora of scientific investigations over the last decades \cite{leibfried2003quantum,bloch2008many,buluta2009quantum,blatt2012quantum}. Over recent years, the field has expanded towards molecules \cite{ni2008high,carr2009cold,krems2009cold,tarbutt2019laser,baum20201d}, as their more complex quantum-state structure opens a broader scope of physics to be studied, including measurements of fundamental constants \cite{chin2009ultracold,acme2014order,cairncross2017precision}, investigations of quantum chemistry \cite{ospelkaus2010quantum,quemener2012ultracold,bohn2017cold}, applications in quantum information \cite{demille2002quantum,krems2009cold,carr2009cold}, and access to novel many-body quantum systems \cite{baranov2012condensed}. Identifying molecules that are relevant for scientific and technological applications and at the same time amenable to laser cooling is an active area of research but, so far, it has almost exclusively focused on neutral molecules \cite{zhelyazkova2014laser,anderegg2017radio,lim2018laser,mcnally2020optical,kozyryev2017sisyphus,zeppenfeld2012sisyphus}.

Molecular ions have broad scientific use cases  \cite{carr2009cold}. Progress on laser cooling and quantum control schemes for molecular ions promises to open new scientific avenues, building on the enormous success of atomic ions in quantum science \cite{monroe1995demonstration,leibfried2003quantum}. For quantum information, the rich internal structure of molecular ions may allow the realization of efficient gate operations and long qubit storage times \cite{demille2002quantum,troiani2011molecular} in the same physical system, akin to neutral molecules \cite{park2017second,hudson2018dipolar,rugango2015sympathetic}. Molecular ions enable Coulomb-mediated two-qubit operations \cite{bruzewicz2019trapped}, and have been proposed as an alternative to neutral molecules \cite{schuster2011cavity, sinhal2022molecular}. They also may allow the realization of qudits, multi-level systems that constitute a powerful extension of the qubit-based quantum information paradigm \cite{moreno2018molecular, albert2020robust}. In addition, cold molecular ions in well-defined quantum states can be employed in collisional studies in quantum chemistry and astrochemistry. For example, gas-phase chemistry in cold interstellar clouds is driven by ion-neutral reactions where the ions are in their lowest electronic, vibrational, and rotational levels due to the relaxation of internal excitations close to the 2.7~K cosmic microwave background \cite{wakelam2012kinetic,mcelroy2013umist}. Fully quantum mechanical calculations for these reactions are beyond current computational capabilities for systems with four or more atoms. Therefore, laboratory measurements with molecules in quantum states similar to those in interstellar space would be helpful to elucidate the chemical kinetics \cite{o2015reaction,de2015merged,tran2018formation,hillenbrand2019experimental,kumar2018low}. 

Today, the standard approach to the preparation of trapped molecular ions relies on sympathetic cooling via a co-trapped ionic \cite{ryjkov2006sympathetic,rugango2015sympathetic} or neutral \cite{hudson2016sympathetic} atomic species. Direct laser cooling of molecular ions has not been explored extensively. Although cooling via co-trapped species has proven effective and useful \cite{sinhal2022molecular}, it is technically challenging. In addition, for sensitive applications in quantum information and precision measurements, where high fidelity and low dephasing are critical \cite{ladd2010quantum}, the presence of a second species may eventually be limiting \cite{nguyen2011challenges,schuster2011cavity}. All-optical cooling schemes should be highly attractive but, so far, only few theoretical studies on the prospects of laser cooling of molecular ions exist \cite{nguyen2011challenges} and most have not accounted for the full rovibrational structure of the ion  under investigation \cite{kang2017ab,li2019laser,chmaisani2021theoretical}. 

\begin{figure}
    \centering
    \includegraphics[width = 0.475\textwidth]{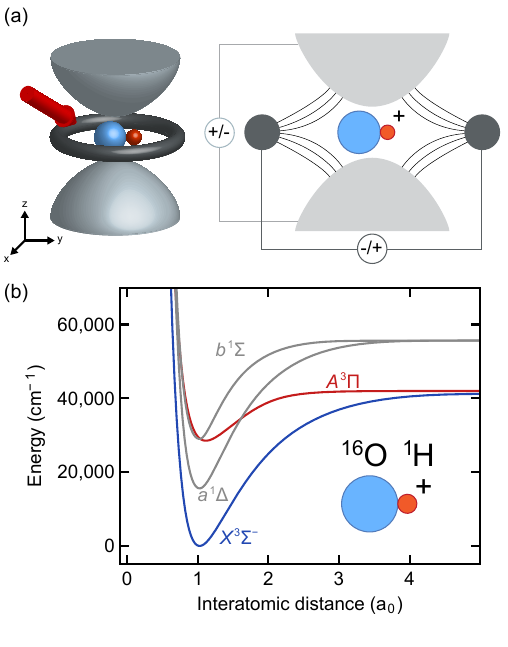}\\
    \caption{Laser cooling of OH$^+$ ions trapped in an ion trap. (a) Schematic of an OH$^+$ ion a cylindrical quadrupole ion trap \cite{march2009quadrupole}. The left image shows a three-dimensional sketch of the ion trap; the right image a cross section, including field lines. The specific trap shown is for illustration purposes only; the cooling scheme is general and can be implemented in other types of ions traps. The cooling and repumping beams are incident collinearly from one direction that has a finite angle with all three Cartesian axes of the trap, providing cooling in three dimensions. (b) Potential energy curves of OH$^+$ \cite{gomez2014OH+}}. Depicted are the low-lying electronic potentials relevant to this study, two with triplet spin character and two with singlet spin character. The inset shows a pictorial representation of an OH$^+$ molecular ion.
    \label{fig:1}
\end{figure}

In this article, we discuss a laser-cooling scheme for the hydroxyl cation, OH$^+$. We envision the scheme to be applied to a single ion or an ensemble of ions held in a deep and tightly confining ion trap, as schematically illustrated in Fig.~\ref{fig:1} (a). In this setting, three-dimensional laser cooling can be achieved with cooling and repumping laser beams incident from a single direction that has a finite angle with all three Cartesian axes of the trap \cite{karl2023laser}. The proposed scheme can provide cooling in the Doppler regime, where the motional energy of the ion is significantly larger than the trap frequency \cite{sinhal2022molecular}. To cool below the Doppler limit, a secondary Raman sideband cooling step may be employed \cite{nguyen2011challenges}. However, in this work we solely focus on the description of the Doppler cooling scheme for OH$^+$.

The OH$^+$ ion is particularly relevant in the context of astrophysics and astrochemistry. The production of cold, pure, and trapped samples of OH$^+$ would enable reaction studies that can help shed light on processes such as the cosmic ray ionization rate of the intersellar medium \cite{van2021water} and the gas-phase pathway to the formation of water \cite{neufeld2017cosmic}. Additional uses of laser-cooled OH$^+$ molecules may be in quantum information. Due to large rotational spacings ($B \sim 500$ GHz \cite{hodges2017fourier}), coherent control of rotational qubits in OH$^+$ would be challenging but possible via a two-photon Raman transfer. Given that Doppler cooling schemes have not been widely discussed for molecular ions, this study also uses OH$^+$ as a proof-of-concept case that may encourage the development of laser cooling schemes for other molecular ions.

The relevant low-lying potential energy curves of OH$^+$ are shown in Fig. \ref{fig:1} (b) \cite{gomez2014OH+}. OH$^+$ has favorable characteristics for laser cooling. Its energy level structure is relatively simple, with its first few electronic states having either a triplet or a singlet nature. Furthermore, OH$^+$ is extremely tightly bound. As can be seen from the minima of the potential energy curves (see Fig. \ref{fig:1}), the bond length is about one $a_0$, where $a_0$ is the Bohr radius. As a result, the vibrational and rotational energy spacings are large, of the order of 3000 cm$^{-1}$ and 60 cm$^{-1}$, respectively\. Due to the large vibrational spacing, only about a dozen vibrational states exist in the $X^3\Sigma^-$ potential below the $A^3\Pi$ potential, which limits the number of decay channels for potential transitions of a laser cooling scheme. However, there is no transition in OH$^+$ with a near-diagonal Franck-Condon factor (FCF), a characteristic that is often believed necessary for a functional cooling scheme \cite{nguyen2011challenges}. The highest branching ratio for a single transition is 0.56 \footnote{This branching ratio is observed between the states $A^3\Pi$ $| 0,1,0 \rangle$ and $X^3\Sigma^-$ $\ket{0,\,0,\,1}$.}. Despite this complication, as we demonstrate below, OH$^+$ supports a laser cooling scheme that is not more complex than state-of-the-art cooling schemes for neutral molecules \cite{vilas2022magneto}. OH$^+$ also has a non-zero nuclear spin of $I=1/2$ \cite{emsley1995elements}, which we do not explicitly take into account in this study. However, we do not expect the resulting hyperfine structure to fundamentally prevent the laser cooling from functioning, as has been seen in earlier demonstrations of laser cooling \cite{shuman2009radiative}.

\section{Laser cycling scheme}
  
\begin{figure}
    \centering
    \includegraphics[width = 0.475\textwidth]{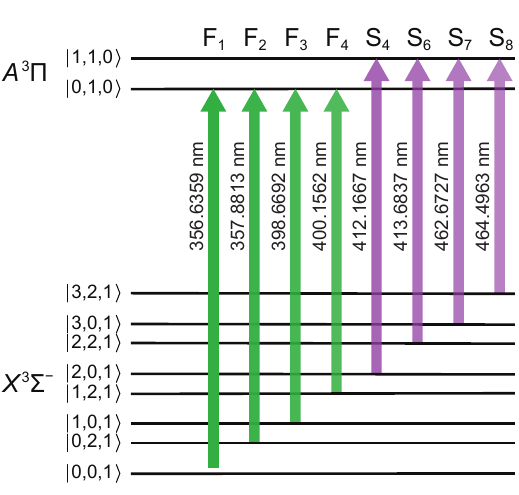}\\
    \caption{Laser cooling scheme. The green (purple) arrows represent the cooling and repumping beams from the $X^3\Sigma^-$ ground electronic state to the $A^3\Pi$ $| 0,1,0 \rangle$ ($| 1,1,0 \rangle$) state. The numbers on the left represent the quantum numbers $|v,N,J\rangle$. Next to each arrow, we indicate the wavelength of each transition. Transitions from left to right are ordered from smaller-to-larger transition wavelength and larger-to-smaller transition strength. The labels above each transition for the first (F) and second (S) legs of the cooling scheme are introduced for easier comparison to Fig. \ref{fig:3}}
    \label{fig:2}
\end{figure}

The proposed cycling scheme makes use of transitions between the electronic ground state of the molecule, $X^3\Sigma^-$, and its first electronically excited triplet state, $A^3\Pi$. The specific transitions of the scheme are shown in Fig.~\ref{fig:2}. We explain below how a sufficient degree of closure can be reached. The required spectroscopic data was extracted from the ExoMol database  \cite{yurchenko2018exomol}, which for OH$^+$ relies on the optical transitions published in Refs.~\cite{hodges2017fourier,bernath2020mollist}. We utilize the provided energy levels, transition frequencies, and Einstein $A$ coefficients.

The logic of the cooling scheme is as follows: We start from the absolute molecular ground state, $X^3\Sigma^-$ $\ket{v = 0,\, N = 0,\, J = 1}$, following Hund's case (b) \cite{hodges2018improved}, where $v$, $N$ and $J$ are the vibrational quantum number, the angular momentum excluding electron and nuclear spin, and the angular momentum excluding nuclear spin, respectively. Due to the $\Delta J = \pm 1$ selection rule, excitation both to $J=0$ and $J=2$ states is possible. Excitation to $J=0$ is favorable as the only decay channel is to $J=1$, reducing the number of needed repumper lasers by a factor of two. In addition, excitation to a low-lying vibrational state is favorable due to higher FCFs. Thus, we choose $A^3\Pi$ $\ket{0,\, 1,\,0}$ as the first excited state of the scheme. The main decays of this state are to the eight $X^3\Sigma^-$ states shown in Fig.~\ref{fig:2}, with only two other observed decays with negligible branching ratios ($\sim 10^{-8}$). Overall, the scattering rate of $A^3\Pi$ $\ket{0,\,1,\,0}$ is relatively small at $3.5\times 10^5$ s$^{-1}$ and by itself would lead to long cooling times. By adding a second excited state to the cycling scheme, the $A^3\Pi$ $\ket{1,\,1,\,0}$ state, it is possible to speed up the cooling time by about a factor of two, as we discuss below. 

\begin{figure}
    \centering
    \includegraphics[width = 0.475\textwidth]{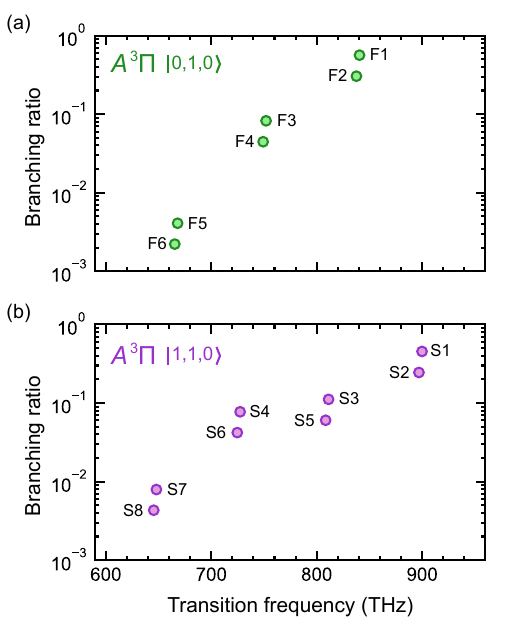}\\
    \caption{Branching ratios for decay from the excited states (a) $A\, ^3\Pi$ $|0,1,0\rangle$ and (b) $A\, ^3\Pi$ $|1,1,0\rangle$. The decay paths for each scheme are numbered in order of decreasing branching ratios and the labels and color coding refer to the transitions shown in Fig. \ref{fig:2}. All decays are to the same states in the $X\, ^1\Sigma^-$ manifold illustrated in Fig. \ref{fig:2}.}
    \label{fig:3}
\end{figure}

Among the decay channels of the excited $A^3\Pi$ $\ket{0, \, 1,\, 0}$ and $A^3\Pi$ $\ket{1, \, 1,\, 0}$ states are 6 and 8 decay channels, respectively, with branching ratios above $10^{-3}$, as shown in Fig.~\ref{fig:3}. This is a practical threshold for the relevance of a transition in a typical laser cooling scheme \cite{bigagli2022laser}. Branching ratios are calculated using the Einstein coefficients $A_{i}$ for spontaneous decay to a state $i$ \cite{yurchenko2018exomol} through the relation $\mathrm{BR}_i = A_{i}/\sum_j A_{j}$, where the sum runs over all possible radiative decay paths to a lower energy state $j$ for a given $A^3\Pi$ state. 

We have investigated the question whether loss to other states could potentially hamper the $X^3\Sigma^- \leftrightarrow A^3\Pi$ cooling scheme. We find that the effects of state mixing between $A^3\Pi$ and the singlet states $b\,^1\Sigma^+$ and $a\, ^1\Delta$ should be minimal. For the nearby $b\,^1\Sigma^+$ state, spin-orbit coupling with a strength of about 75 cm$^{-1}$ \cite{merer1975ultraviolet} leads to an admixture of 0.5$\%$ $b\,^1\Sigma^+$ $|v = 0\rangle$ to the $A\,^3\Pi$ $|v = 0,1\rangle$ states, calculated from the diagonalization of the Hamiltonian of these states with an off-diagonal spin-orbit contribution. Although this is a non-negligible admixture, it is not expected to lead to relevant losses out of the cycling scheme (see Appendix). This is due to the unusual structure of OH$^+$, where decay from $b\,^1\Sigma^+$ to $a\,^1\Delta$ is suppressed by the $\Delta \Lambda =0, \pm1$ selection rules \cite{bernath2020spectra}. Furthermore, direct coupling between $b\,^1\Sigma^+$ and vibrationally excited $a\,^1\Delta$ states should not be an issue, as the only loss channel via mixing of $a\,^1\Delta$ can be due to relaxation into lower vibrational states of $a\,^1\Delta$. However, spontaneous decay between vibrational states is suppressed in the first order \cite{bernath2020spectra} and driving of such transitions by black-body radiation would be slow compared to  the timescale of the cooling scheme. In line with these arguments, to our knowledge, such decays in OH$^+$ have not been reported. Finally, we note that loss from predissociation \cite{sun2023probing} is fully suppressed for the $X^3\Sigma^- \leftrightarrow A^3\Pi$ cooling scheme.

\section{Results \& Discussion}

Using the branching ratios from Fig. \ref{fig:3}, we quantify the closure that can be achieved by adding an increasing number of repumping transitions. We calculate the number of scatterings that will lead to a probability of $10\%$ ($n_{10\%}$) and $90\%$ ($n_{90\%}$) for retaining an ion in the cycling scheme \cite{di2004laser}. calculations are made assuming the use of two excited states, as shown in Fig. \ref{fig:2}. We employ a Monte Carlo model in which an ion is initialized in the ground state and then each scattering event is simulated by updating the probability of populating each state in the cooling scheme after each step. Table \ref{tab:ClosureScatterings} shows the results of these calculations. The quantity $p$ is the closure of the scheme, calculated via $n_{x\%} = \ln(x/100)/\ln(p)$ \cite{di2004laser}. A conservative estimate of the time to complete $n_{x\%}$ scattering events, $t_{x\%}$, is also provided. To calculate this quantity we use the relation $t_{x\%} = n_{x\%}/R$, where $R = \Gamma/(G + 1 + 2 \sum I_{\text{sat},i}/I_i )$ is the scattering rate \cite{tarbutt2013design, FITCH2021157}. Here, $\Gamma$ is the excited state linewidth; $G$ is the number of driven transitions; $I_{\text{sat},i} = \pi h c \Gamma / 3 \lambda_i^3$ is the saturation intensity \cite{FITCH2021157} of the $i^\text{th}$ transition; and $I_i$ is the intensity of the laser addressing said transition, set to $I_i = 10^3\,$mW$\,$cm$^{-2}$, which is an intensity that is easily achievable in an experiment at the given wavelengths. In the definition of the saturation intensity, $c$ is the speed of light, and $\lambda_i$ the wavelength of the addressed transition. This results in an overestimate of $t_{x\%}$ as the expression for $R$ assumes a single excited state, only $A^3\Pi$ $\ket{ 0, \, 1,\, 0}$. In an experiment, where several repumping transitions cycle through a second excited state, here $A^3\Pi$ $\ket{ 1, \, 1,\, 0}$, the scattering rate is boosted and the cooling time lowered, as has been experimentally demonstrated in laser cooling schemes for neutral molecules, e.g.~in Ref.~\cite{kozyryev2017sisyphus}.

\begin{table}
    \centering
    \caption{Closure, number of scatterings, and scattering time for an increasing number of repumping transitions.}
    \begin{tabular}{ccccccc} 
    \hline
    \hline
    Driven transitions & $p$ & $n_{10\%}$ & $t_{10\%}$ (ms) & $n_{90\%}$ & $t_{90\%}$ (ms) \\ \hline 
     F1 & 0.631 & 5 & 0.02 & 1 &  \\
     F1-F2 & 0.873 & 17 & 0.1  & 1 &  \\
     F1-F3 & 0.950 & 45 & 0.5  & 3 & 0.02  \\
     F1-F4 & 0.994 & 361 & 5 & 17 & 0.2 \\
     F1-F4, S4 & 0.997 & 918 & 17 & 43 & 0.8 \\
     F1-F4, S4, S6 & 0.9986 & 15$\,$941 & 800 & 731 & 36 \\
     F1-F4, S4, S6-S7 & 0.99995 & 44$\,$287 & $2.6\times10^3$ & 2$\,$028 & 100 \\
     F1-F4, S4, S6-S8 & 0.9999997 & 2$\,$563$\,$381 & $1.9\times10^6$ & 117$\,$295 & $9.0\times10^4$ \\
    \hline
    \hline
    \end{tabular}
    \label{tab:ClosureScatterings}
\end{table}

\begin{table}
    \centering
    \caption{Results of the photon scatterings analysis starting at cryogenic and room temperature.} 
   \begin{tabular}{cc|cc} 
    \hline
    \hline
      \multicolumn{2}{c|}{$T = 4$ K} & \multicolumn{2}{c}{$T = 300$ K}\\
    $n_{\mathrm{cool}}$ &  $t_{\text{cool}}$ (ms) & $n_{\mathrm{cool}}$ &  $t_{\text{cool}}$ (ms) \\ \hline
     $2.3 \times 10^3$ & 45 & $2.0 \times 10^4$ & 400 \\ 
    \hline
    \hline
    \end{tabular}
    \label{tab:Cooling}
\end{table}

We calculate the number of scatterings, $n_{\rm cool}$, required to bring the sample to close-to-zero motional temperature from room (300 K) and cryogenic (4 K) temperatures. In addition, we calculate the time $t_{\rm cool}$ necessary for this process. For the experimental setup, we assume that the ion is electrostatically trapped and that cooling and repumping laser beams come from a single direction that has a finite angle with all three Cartesian trap axes, as illustrated in Fig.~\ref{fig:1} (a). Using a single incoming direction for the lasers enables cooling in all spatial directions thanks to the confinement provided by the ion trap, as suggested in Ref.~\cite{karl2023laser}, and greatly simplifies the laser setup. Table \ref{tab:Cooling} summarizes the results. Cooling OH$^+$ ions down from room temperature, will require $2 \times 10^4$ scattering events. Comparing this requirement to the results in Table \ref{tab:ClosureScatterings}, this can be achieved with 90$\%$ efficiency within $<500$ ms using 8 transitions (F1-F4, S4, and S6-S8). Cooling OH$^+$ ions from cryogenic temperatures, will require $2.3 \times 10^3$ scattering events, which can be achieved with 90$\%$ efficiency within $<50$ ms using 6 transitions (F1-F4, S4, and S6). We note that OH$^+$ can be easily produced at cryogenic temperatures, as demonstrated in earlier work via electronic impact on water vapor \cite{kumar2018low,tran2018formation}. It is important to note that in practical experiments less stringent requirements on the probability of retention should be acceptable. For a $10\%$ probability of retention, cooling from room temperature can be achieved with 6 to 7 lasers (F1-F4, S4, S6-S7) within about 1 s, and from cryogenic temperatures with 5 to 6 lasers  (F1-F4, S4, S6) within about 50 ms. Hence, on average it would take 10 s and 0.5 s, respectively, to successfully initialize the system, a time that is short compared to the storage time of a molecular ion once it is trapped and cooled.     

\section{Conclusions}

In summary, we have presented a direct laser cooling scheme of OH$^+$. The scheme is expected to work both for a single trapped ion and trapped ensembles. In the case of an ensemble, the direction of the laser cooling beams could even be arbitrary, as the Coulomb interactions in the ensemble are expected to couple all directions of motion \cite{sinhal2022molecular}. While the scheme requires some technical effort, laser cooling of OH$^+$ would come with several benefits: Cold samples can be prepared without the need for sympathetic cooling via neutral or ionic atoms, and efficiently temperatures can be reached where Raman sideband cooling can take over to cool ions to the vibrational ground state (of the trap). Rapid progress in laser technology \cite{corato2023widely} promises to drastically reduce the technical complications involved in a cooling scheme with multiple lasers. Also, due to the confined trapping region of an ion trap, the lasers can be tightly focused and we expect that only very moderate laser powers are be necessary. Once cooled, OH$^+$ will enable studies relevant for gas-phase astrochemistry of interstellar clouds. Furthermore, quantum control of OH$^+$ will be an enabling step for its use as a qubit or qudit in quantum information experiments. Finally, this work, similar to the case of C$_2$ which we discussed in earlier work \cite{bigagli2022laser}, suggests that also molecules with non-diagonal FCFs can be amenable to laser cooling. We hope this work will inspire further investigations on the application of laser cooling techniques to more complex molecules that are scientifically relevant, but have so far been deemed unsuitable for laser cooling.  

\hspace{12pt}

\section{Acknowledgements}

We thank Peter F.\ Bernath, Tim de Jongh, \'{A}bel K\'{a}losi, Ian Stevenson, and Sergey Yurchenko for stimulating discussions, and Octavio Roncero for help with the potential energy curves. This work was supported by a Columbia University Research Initiative in Science and Engineering (RISE) award. D.W.~Savin was additionally supported by the NASA Astrophysics Research and Analysis program under 80NSSC19K0969. S.W.~acknowledges additional support from the Alfred P.~Sloan Foundation.

\section{Appendix}

As noted in the main text, a loss channel for the $X^3\Sigma^- \leftrightarrow A^3\Pi$ cooling scheme could arise from spin-orbit coupling between $A^3\Pi$ and nearby singlet states. Here, we provide more detail to show that such processes should be sufficiently suppressed for practical purposes. As mentioned in the main text, the $0.5\%$ admixture of $b^1\Sigma^+ \, |v=0 \rangle$ to the $A^3\Pi\, |v=0,1\rangle$ states does not open up a loss channel to first order due to prohibitive selection rules. To second order, decay could happen via the admixture of $^1\Pi$ character to the $b^1\Sigma$ potential, with the closest $^1\Pi$ state lying about 15,000 cm$^{-1}$ above the $b^1\Sigma$ potential \cite{gomez2014OH+}. With a coupling constant of 75 cm$^{-1}$ (assuming a coupling similar to $b^1\Sigma^+$ and $A^3\Pi$; to our knowledge there is no reported literature value) this leads to a $10^{-5}$ admixture of $^1\Pi$ character to the $b^1\Sigma$ state. Taken together, this amounts to a loss channel on the $5\times10^{-8}$ level. Loss channels allowed through higher order processes should contribute significantly less. Also direct coupling between $A^3\Pi$ and $a^1\Delta$ is not expected to lead to significant loss. The only decay path out of the cycling scheme would be spontaneous decay into lower lying vibrational states of $a^1\Delta$ which is forbidden \cite{bernath2020spectra}. Based on these arguments it is highly probable that the proposed cooling scheme will be sufficiently closed, especially taking into account that only closure to the $10^{-4}$ level will be needed for the scheme to be practically useful (see main text and Table \ref{tab:Cooling}).

\end{document}